\documentclass[prc,byrevtex,showpacs,aps,twocolumn]{revtex4-1}
\usepackage{epsfig,graphicx,float,pst-all, rotating}
\usepackage{hyperref}
\usepackage{amssymb}
\usepackage{amsmath}
\usepackage{multirow}
\usepackage{color}  

\newcommand{\be}{\begin{equation}} 
\newcommand{\ee}{\end{equation}}
\newcommand{\bea}{\begin{eqnarray}} 
\newcommand{\eea}{\end{eqnarray}}
\newcommand{\bc}{\begin{center}}
\newcommand{\ec}{\end{center}}

\begin{document}

\title{Enhanced subbarrier fusion for proton halo nuclei} 

\author{Raj Kumar}

\author{J. A. Lay}

\author{A. Vitturi}

\affiliation{ Dipartimento di Fisica e Astronomia ``Galileo Galilei", Universit\`{a} di Padova, via Marzolo, 8, I-35131 Padova, Italy}
\affiliation{INFN, Sezione di Padova, via Marzolo, 8, I-35131 Padova, Italy}

\begin{abstract}
In this short note we use a simple model to describe the dynamical effects of break-up processes in the subbarrier fusion involving weakly bound nuclei.  We model two similar cases involving either a neutron or a proton halo nucleus, both schematically coupled to the break-up channels.  We find that the decrease of the coulomb barrier in the proton break-up channel leads, {\it ceteris paribus}, to a larger enhancement of the subbarier fusion probabilities with respect to the neutron-halo case.

\end{abstract}

\maketitle

Subbarrier heavy-ion fusion processes have been in the last decades an interesting issue for the low-energy nuclear physics community for the natural link involved between structure and dynamics.  It has been in fact recognized that the basis feature characterizing the subbarrier behavior is the dynamical coupling to the internal degrees of freedom of the two fusing partners~\cite{Bal98,Das83b,Hag12}.   The proper description of a fusion process, therefore, is essentially demanding to single out the relevant coupled channels involved and to determine the associated diagonal and coupling potentials.  This makes the situation with weakly bound nuclei more complex, due to the non trivial inclusion of the strongly coupled continuum break-up channels and the consequent opening of final three-body (or four, in the case of two-particle halo nuclei) channels.  This had led from a theoretical point of view of diverging results on the enhancement/suppression of the fusion probabilities, and to extremely difficult experimental measurements to determine (and separate) different fusion and reaction channels~\cite{Agu11,Agu09,Scu11,Hag00,Vin13,Gom11,Nak04}.

Given the complexity of the situation, every case behaves differently and has to be specifically treated, with particular ion-ion potentials, associated heights of the coulomb barrier, coupling form factors, specific relevant transfer channels and $Q$-values.  For this reason it is not easy, in a fully treated coupled-channel description, to single out the role of specific issues.  One of these is the possible role of the charged break-up channels in proton-halo nuclei with respect to the more common neutron break-up channels in neutron-halo nuclei.   For this reason we introduce here a very simplified two-channel model, the first being the entrance channel and the second representing the full set of continuum break-up channels.  In this channel we neglect the ejected particle (neutron or proton) and properly rescale energies and ion-ion potential.  Our model has been applied, as representative cases of neutron or proton haloes, to the fusion with $^{58}$Ni of either $^{11}$Be and $^{8}$B.  To single out just the dynamical effects due to the neutron/proton nature of the two halo nuclei, potentials in the different channels have been constructed using the simple parameterization of Broglia and Winther~\cite{BW} and an equal strength for the coupling between entrance channels and the "break-up" ones.

\begin{figure}[h]
\includegraphics[width=1.\columnwidth,clip=true]{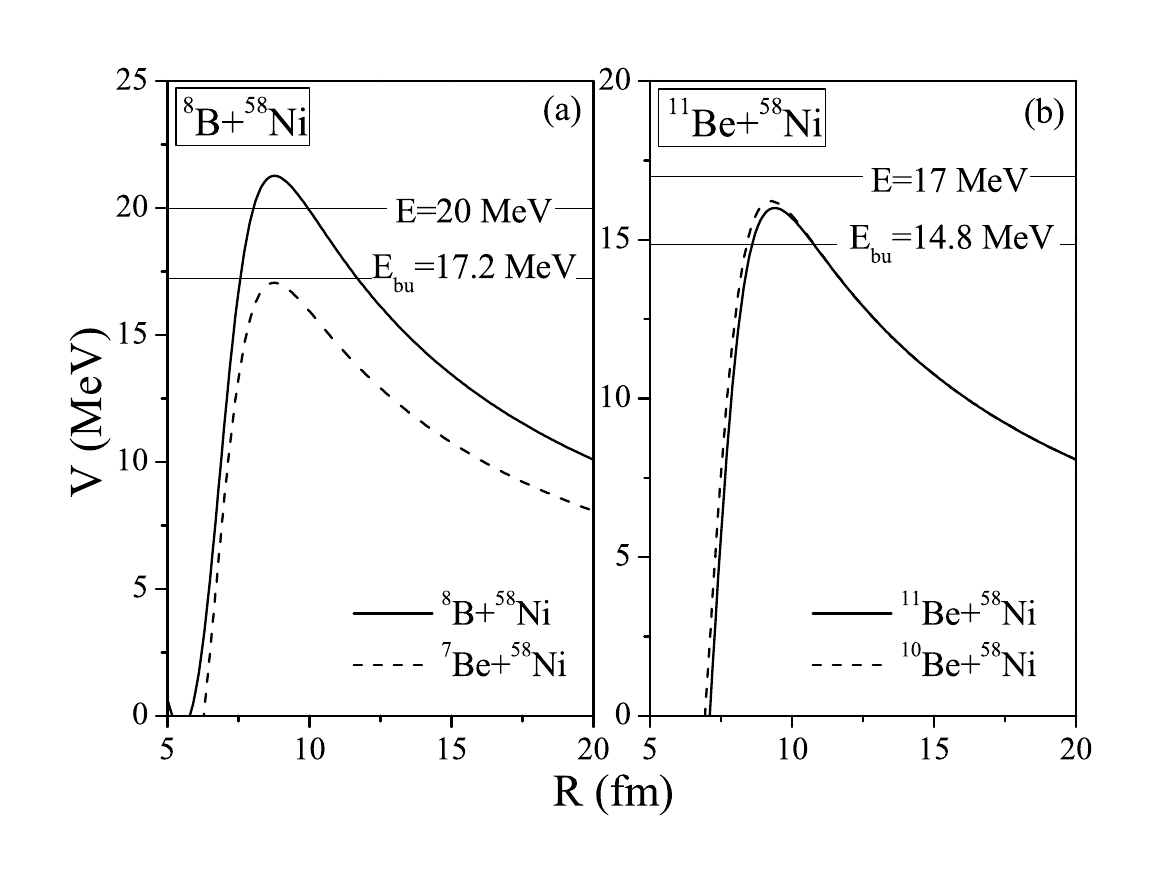}
\caption{Ion-ion potentials for $^{8}$B+$^{58}$Ni in the left frame and $^{11}$Be+$^{58}$Ni in the right frame (solid lines). The dashed lines corresponds to the break-up channels, i.e. for the $^{7}$Be+$^{58}$Ni   and $^{10}$Be+$^{58}$Ni respectively. The nuclear part of the potential is computed according the proximity potential of Broglia and Winther~\cite{BW}. }
\label{pot}
\end{figure}

In Fig.~\ref{pot} we display the resulting ion-ion potentials for the $^{8}$B+$^{58}$Ni (left frame) and $^{11}$Be+$^{58}$Ni (right frame) reactions.  For comparison in the same figures we also display the corresponding ion-ion potentials in our ''break-up" channels, i.e. for the $^{7}$Be+$^{58}$Ni   and $^{10}$Be+$^{58}$Ni cases.  For a quick view, we also show in the figure as a line one  energy $E$ in the incoming channel (20~MeV in the case of $^{8}$B and 17~MeV in the case of $^{11}$Be) and the corresponding energy in the break-up channel. This energy can be estimated by subtracting the energy needed for break-up and the average excitation energy, $\langle E^* \rangle$, in the core-nucleon relative motion, and then sharing the energy between then according to a distant break-up scenario. In this way, we consider $E_{bu}=(E-S_{1N}-\langle E^* \rangle) \cdot \frac{A-1}{A}$. $S_{1N}$ reads for the one neutron or one proton separation energy, i.e. $S_{1p}=0.136$~MeV for $^8$B and $S_{1n}=0.504$~MeV for $^{11}$Be. $\langle E^* \rangle$ is approximated by the peak energy for the dipole electromagnetic transition probabilities, $\langle E^* \rangle=0.5$~MeV for $^8$B and $\langle E^* \rangle=0.4$~MeV for $^{11}$Be. 
It is evident from the figure that while in the neutron case the barriers in the incoming and break-up channels are similar (while the energy at disposal in the latter is smaller), in the proton case the reduction in energy in the break-up channel is more than compensated by the lower Coulomb barrier due to the reduced charge in the projectile.

Fusion probabilities are calculated by solving the corresponding coupled-channel equations under ingoing-wave boundary conditions (IWBC). The coupled-channel formalism for direct reaction processes given by Austern~\cite{Aus87} expands the total wave function in terms of the wavefunction for the internal state of the projectile $\phi_{\beta}$ and the radial wave functions $\chi_{\beta}$ that acounts for the relative motion between projectile and target:
\begin{equation}
\Psi^{(+)}=\Sigma_{\beta}\frac{\chi_{\beta}(R)}{R}\phi_{\beta}.  
\label{eq.1}
\end{equation}
This leads to a set of coupled equations for the radial wave functions:
\begin{equation}
\frac{d^{2}\chi_{\beta}}{dR^{2}}+\frac{2\mu_{\beta}}{\hbar^{2}}[E_{\beta}-V_{\beta}^{eff}(R)]\chi_{\beta}
=\frac{2\mu_{\beta}}{\hbar^{2}}\Sigma_{\alpha\ne\beta}V_{\beta\alpha}^{coup}(R)\chi_{\alpha}  
\label{eq.3}
\end{equation}
In these expression 
V is the interaction potential while, for a given channel $\beta$, $\mu_{\beta}$ is the reduced mass, 
and $E_{\beta}$ is the relative energy.



In our model case, we will only consider two channels, the incomming channel and one channel representative of the break-up and later fusion without the ejected particle.
The two channel problem in one spatial dimension $R$ is given by:
\begin{eqnarray}
&&\frac{d^{2}\chi_{1}}{dR^{2}}+\frac{2\mu_{1}}{\hbar^{2}}[E_{1}-V_{1}]\chi_{1}
=\frac{2\mu_{1}}{\hbar^{2}}V_{coup}\chi_{2},\nonumber\\
&&\frac{d^{2}\chi_{2}}{dR^{2}}+\frac{2\mu_{2}}{\hbar^{2}}[E_{2}-V_{2}]\chi_{2}
=\frac{2\mu_{2}}{\hbar^{2}}V_{coup}\chi_{1},
\label{eq.5}
\end{eqnarray}
where, in our case, $E_{1}=E$, the incoming energy, and $E_{2}=E_{bu}$, the enegy in the break-up channel.

The total potential for each channel $V_{1,2}(R)$ is given by the sum of Coulomb and a nuclear proximity potential given by Broglia and Winther~\cite{BW} parameterization. The coupling potential $V_{coup}$ is given as a derivative Woods Saxon form with same radius and difuseness of the proximity potential for the incoming channel. The strength is set to a 10\% of the strength of the same proximity potential.



The coupled channel equations are solved by imposing the boundary conditions that there are only incoming waves at R=$R_{min}$, i.e. the minimum position of the Coulomb pocket inside the barrier, and there are only outgoing waves at infinity for all channels except for the entrance channel ($\beta$=1), which has an incoming wave with amplitude one as well. This boundary condition is referred to as the incoming wave boundary condition (IWBC)~\cite{Bal98,Hag12,Das83}, and is valid for heavy-ion reactions, where there is strong absorption inside the Coulomb barrier. 
The numerical solution is matched to a linear combination of incoming and outgoing and Coulomb wave functions at finite distance $R_{max}$ beyond which both the nuclear proximity and the 
coupling potential are negligible.
The boundary condition of a wave incident from the right in channel $\beta$=1 and transmitted and reflected waves in both channels is given by,
\begin{eqnarray}
\chi_{\beta}(R) \xrightarrow{R\rightarrow\infty} & \delta_{\beta1} H^{(-)}_{\ell}(k_{\beta}R)&+~~r_{\beta}H^{(+)}_{\ell}(k_{\beta}R); \nonumber\\
\chi_{\beta}(R=R_{min}) =&t_{\beta} H^{(-)}_{\ell}(k_{\beta}R),&
\label{eq:12}
\end{eqnarray}
where $\ell$ is angular momentum, $H^{(+)}_{\ell}$ and $H^{(-)}_{\ell}$ are the outgoing and incoming Coulomb wave functions, respectively and
$k=\sqrt{2\mu E/\hbar^2}$ is the wave number associated with the energy $E$.
The total transmission probability is then given by,
\begin{eqnarray}
T=\sum_{\beta}\mid T_{\beta}^2\mid = |t_1|^2+\frac{v_2}{v_1} |t_2|^2
\label{eq:13}
\end{eqnarray}
where $v_1$ and $v_2$ are the velocities corresponding to channel 1 and 2.

The fusion cross-section, in terms of partial waves, is given by
\begin{equation}
\sigma=\sum_{\ell=0}^{\ell_{max}}\sigma_{\ell}=\frac{\pi\hbar^2}{2\mu_{1} E}\sum_{\ell=0}^{\ell_{max}}(2\ell+1)T_{\ell}(E).
\label{eq:14}
\end{equation}

The probability of transmission for the partial wave can also be calculated simply by a shift of energy,
\begin{equation}
T_{\ell}\cong T_{0}\left[E-\frac{\ell(\ell+1)\hbar^{2}}{2\mu_{1} r_{0}^2} \right],
\label{eq:15}
\end{equation}
where $r_{0}$ is the position of the barrier for the s-wave~\cite{Bal98}.

\begin{figure}
\includegraphics[width=1.0\columnwidth,clip=true]{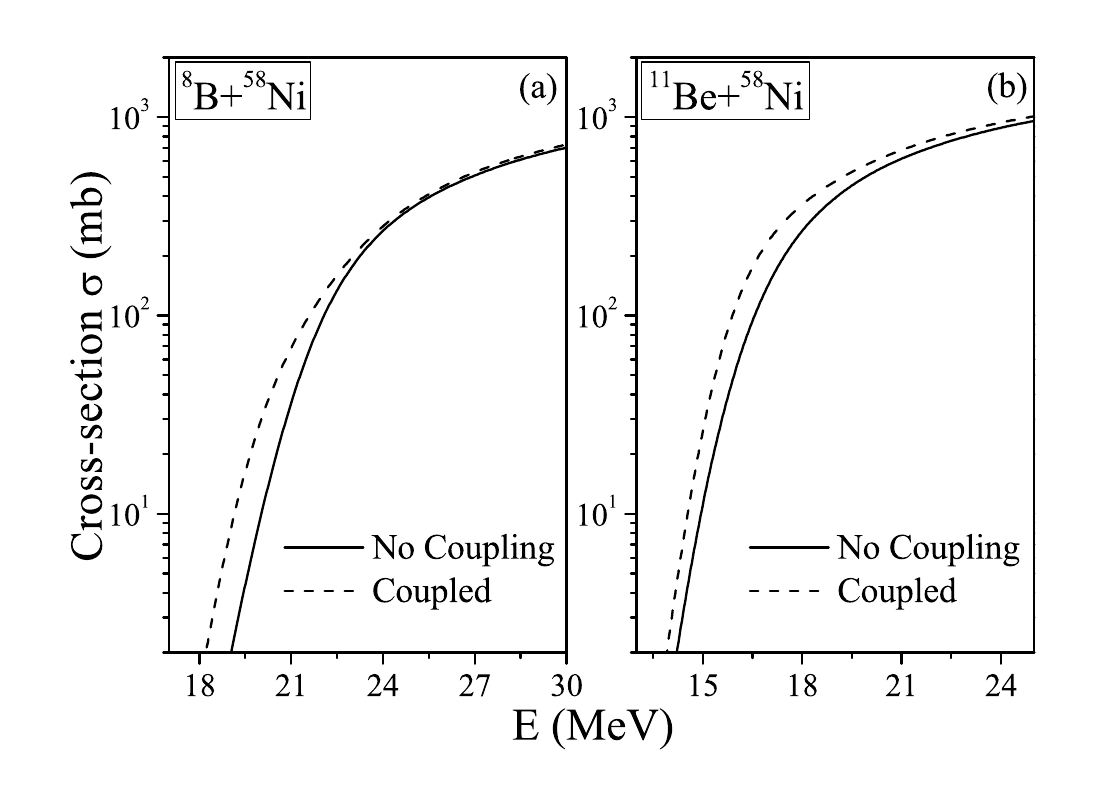}
\caption{Fusion cross sections for the $^{8}$B+$^{58}$Ni (left panel) and $^{11}$Be+$^{58}$Ni (right panel) reactions. Solid lines represent the case without break-up, with a single channel and no coupling, whereas the dashed lines show the two channels case with coupling to the proton (left) and neutron (right) break-up channels.} 
\label{fullxsec}
\end{figure}

\begin{figure}
\includegraphics[width=0.8\columnwidth]{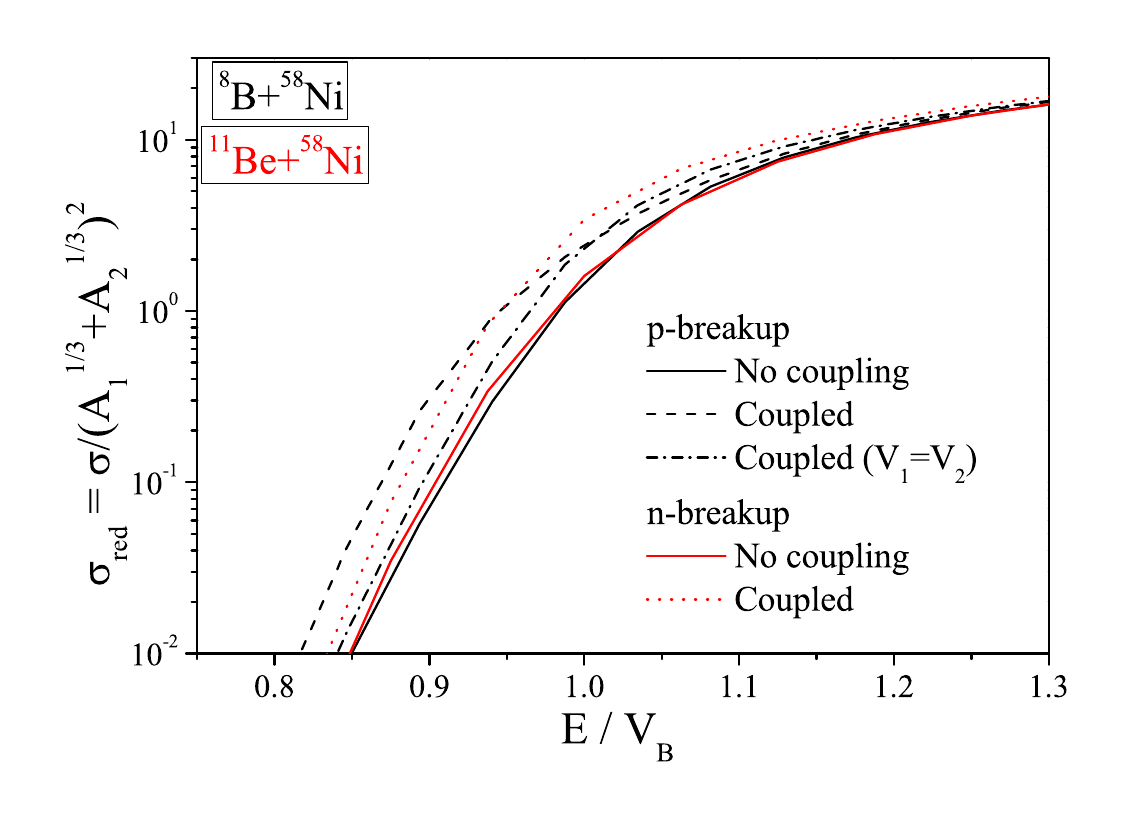}
\caption{(Color online) Cross section divided by the square of the interaction radius versus the energy divided by the estimation of the Coulomb barrier in the incoming channel ($V_{B}$) for $^{8}$B+$^{58}$Ni and $^{11}$Be+$^{58}$Ni fusion reactions. We compare the no coupling cases for both reactions (solid line) with the proton (dotted line) and neutron (dashed line) break-up cases.}
\label{redxsec}
\end{figure}

The resulting cross section for both $^{8}$B+$^{58}$Ni and $^{11}$Be+$^{58}$Ni fusion reactions are shown in Fig.~\ref{fullxsec}. For each reaction, we compare the situation without break-up, where there is no coupling to the second channel (solid lines), with the possibility of coupling to the break-up channel (dashed lines). In both cases, and as a result of this coupling, a certain enhancement is found. In order to compare both cases appropriately, we show in Fig.~\ref{redxsec} a reduced fusion cross sections in terms of the collision radius of each reaction versus the energy divided by the estimated Coulomb barrier. As expected, the two no-coupling cross sections coincide almost perfectly, whereas the coupling cases show different results. Here, it is clearly seen how the proton break-up case has a larger cross section at low energies. On the other hand, the neutron break-up case has a larger enhancement at energies inmediately close to the energy of the Coulomb barrier. For the sake of cancelling the effects of choosing two different nuclei for the neutron and the proton case we add a third case in Fig.~\ref{redxsec} for the $^{8}$B+$^{58}$Ni case where the same potential, and so the same Coulomb barrier, is used for both channels, $V_{2}=V_{1}$ (dot-dashed line). This case is similar to consider that the $^{8}$B looses one neutron instead of a proton. As expected, the cross section follows the same trend as the $^{11}$Be+$^{58}$Ni but with an apparently smaller enhancement.

\begin{figure}
\includegraphics[width=0.8\columnwidth,clip=true]{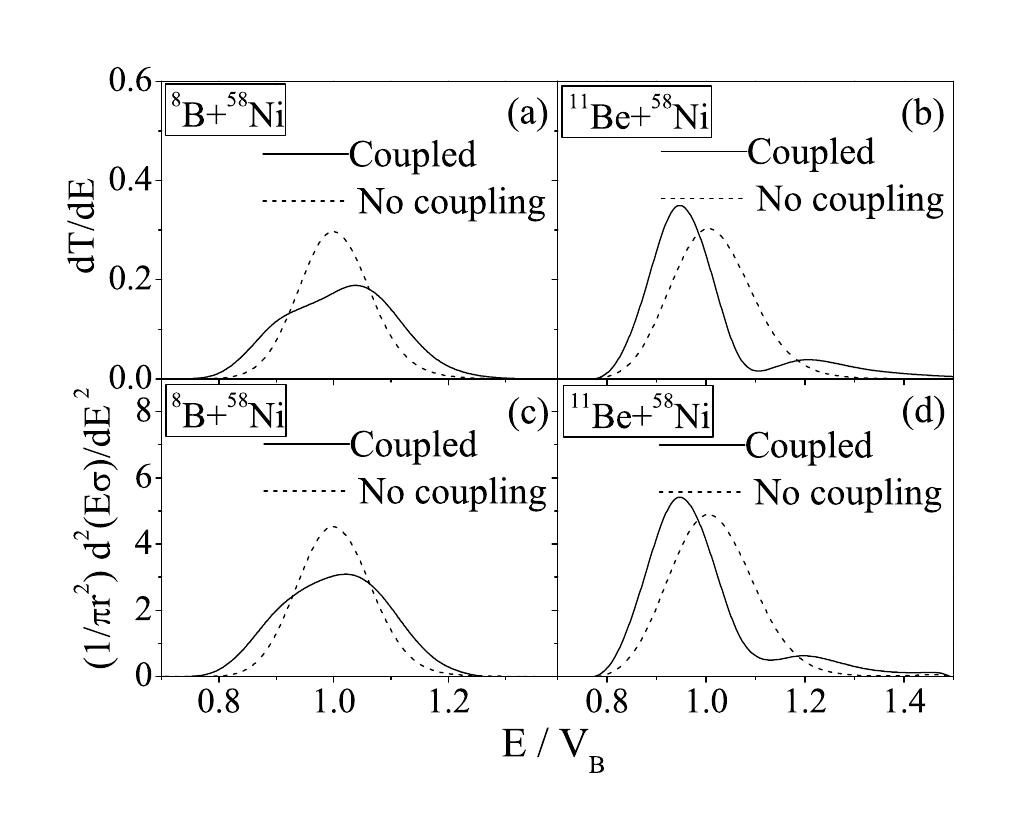}
\caption{Barrier distributions for the $^{8}$B+$^{58}$Ni (left panels) and $^{11}$Be+$^{58}$Ni (right panels) fusion reactions both with (dashed) and without (solid) coupling to the break-up channel. In upper panels we show the derivative of the transmission factor for $\ell=0$ whereas in the lower panels we evaluate the second derivative of the fusion cross section times the energy.}
\label{bardist}
\end{figure}

In order to clarify which processes are giving rise to these two different behaviors, it is useful to show the barrier distributions for both reactions. This can be done by evaluating the second energy derivative of the product of the cross section and the energy, or the first derivative of the transmission for $\ell=0$. Both observables are shown in Fig~\ref{bardist}. A clear difference between the proton and neutron induced effects on fusion is found. Both cases present two barriers as expected according to Fig.~\ref{pot}. However, in the proton case, the secondary barrier is below the barrier in the incoming channel and so it allows a larger enhancement at low energies. Instead, in the neutron case, the secondary barrier is at a higher energy. Therefore, the neutron enhancement simply arises from the displacement towards a lower energy of the final effective Coulomb barrier.

These results obtained here are similar to the effect of negative or positive $Q$-values on barrier penetration~\cite{Das97,Das83b}. As shown, for example, in figure 5.1 in~\cite{Das97}, the positive $Q$-value case shows the same cross section and barrier distribution as the proton break-up case, and the same parallelism is found for negative $Q$-value and neutron break-up cases. Indeed, effective $Q$-values can be considered and compared from the difference between the energies and the barriers in each channel. This effective $Q$-value may be evaluated as
\begin{equation}
 Q_{eff}=(E_{bu}-V_{B}^{2})-(E-V_{B}^{1}),
\end{equation} 
where $V_{B}^{1}$ and  $V_{B}^{2}$ are the energies of the Coulomb barrier for the incoming and break-up channels respectively. Here we have also neglected the effect of the separation energy and the average excitation energy of the projectile. Looking at the energies plotted in Fig.~\ref{pot}, we obtain $Q_{eff}=1.97$~MeV for the proton case and $Q_{eff}=-1.12$~MeV for the neutron case.

The exact value for $Q_{eff}$ will depend on the incoming energy. Nevertheless, it can be shown that it is always negative for the neutron case, whereas it is positive for the proton case at energies around or bellow the Coulomb barrier. Therefore, the differences between the energies and the Coulomb barrier due to the loss of a neutron or a proton can explain the results obtained in both cases.

In conclusion, the possibility of proton break-up produces an enhancement of the subbarrier fusion. Similar results were also found by Nakatsukasa \textit{et al.}~\cite{Nak04} in a time-dependant approach. This fact can explain the enhancement recently found for the proton halo nucleus $^{8}$B~\cite{Agu11}.  This enhancement is larger than in the neutron case, and also the energy distribution is far different. Indeed, for the neutron case, the enhancement is mainly due to a displacement in the energy of the Coulomb barrier.   This can also explain why it is unclear if neutron halo produces or not an enhanced subbarrier fusion.

\begin{acknowledgments}
 This work has been supported by MIUR research fund PRIN 2009TWL3MX. The authors acknowledge L.~F.~Canto for useful discussions.

\end{acknowledgments}

\bibliography{fusion}

\begin{thebibliography}{14}%
\makeatletter
\providecommand \@ifxundefined [1]{%
 \@ifx{#1\undefined}
}%
\providecommand \@ifnum [1]{%
 \ifnum #1\expandafter \@firstoftwo
 \else \expandafter \@secondoftwo
 \fi
}%
\providecommand \@ifx [1]{%
 \ifx #1\expandafter \@firstoftwo
 \else \expandafter \@secondoftwo
 \fi
}%
\providecommand \natexlab [1]{#1}%
\providecommand \enquote  [1]{``#1''}%
\providecommand \bibnamefont  [1]{#1}%
\providecommand \bibfnamefont [1]{#1}%
\providecommand \citenamefont [1]{#1}%
\providecommand \href@noop [0]{\@secondoftwo}%
\providecommand \href [0]{\begingroup \@sanitize@url \@href}%
\providecommand \@href[1]{\@@startlink{#1}\@@href}%
\providecommand \@@href[1]{\endgroup#1\@@endlink}%
\providecommand \@sanitize@url [0]{\catcode `\\12\catcode `\$12\catcode
  `\&12\catcode `\#12\catcode `\^12\catcode `\_12\catcode `\%12\relax}%
\providecommand \@@startlink[1]{}%
\providecommand \@@endlink[0]{}%
\providecommand \url  [0]{\begingroup\@sanitize@url \@url }%
\providecommand \@url [1]{\endgroup\@href {#1}{\urlprefix }}%
\providecommand \urlprefix  [0]{URL }%
\providecommand \Eprint [0]{\href }%
\@ifxundefined \urlstyle {%
  \providecommand \doi  [0]{\begingroup \@sanitize@url \@doi}%
  \providecommand \@doi [1]{\endgroup \@@startlink {\doibase
  #1}doi:\discretionary {}{}{}#1\@@endlink }%
}{%
  \providecommand \doi  [0]{doi:\discretionary{}{}{}\begingroup
  \urlstyle{rm}\Url }%
}%
\providecommand \doibase [0]{http://dx.doi.org/}%
\providecommand \Doi [0]{\begingroup \@sanitize@url \@Doi }%
\providecommand \@Doi  [1]{\endgroup\@@startlink{\doibase#1}\@@Doi}%
\providecommand \@@Doi [1]{#1\@@endlink}%
\providecommand \selectlanguage [0]{\@gobble}%
\providecommand \bibinfo  [0]{\@secondoftwo}%
\providecommand \bibfield  [0]{\@secondoftwo}%
\providecommand \translation [1]{[#1]}%
\providecommand \BibitemOpen [0]{}%
\providecommand \bibitemStop [0]{}%
\providecommand \bibitemNoStop [0]{.\EOS\space}%
\providecommand \EOS [0]{\spacefactor3000\relax}%
\providecommand \BibitemShut  [1]{\csname bibitem#1\endcsname}%
\bibitem [{\citenamefont {Balantekin}\ and\ \citenamefont
  {Takigawa}(1998)}]{Bal98}%
  \BibitemOpen
  \bibfield  {author} {\bibinfo {author} {\bibfnamefont {A.~B.}\ \bibnamefont
  {Balantekin}}\ and\ \bibinfo {author} {\bibfnamefont {N.}~\bibnamefont
  {Takigawa}},\ }\Doi {10.1103/RevModPhys.70.77} {\bibfield  {journal}
  {\bibinfo  {journal} {Rev. Mod. Phys.},\ }\textbf {\bibinfo {volume} {70}},\
  \bibinfo {pages} {77} (\bibinfo {year} {1998})}\BibitemShut {NoStop}%
\bibitem [{\citenamefont {Dasso}\ \emph
  {et~al.}(1983){\natexlab{a}}\citenamefont {Dasso}, \citenamefont {Landowne},\
  and\ \citenamefont {Winther}}]{Das83b}%
  \BibitemOpen
  \bibfield  {author} {\bibinfo {author} {\bibfnamefont {C.~H.}\ \bibnamefont
  {Dasso}}, \bibinfo {author} {\bibfnamefont {S.}~\bibnamefont {Landowne}}, \
  and\ \bibinfo {author} {\bibfnamefont {A.}~\bibnamefont {Winther}},\ }\Doi
  {10.1016/0375-9474(83)90316-0} {\bibfield  {journal} {\bibinfo  {journal}
  {Nucl. Phys. A},\ }\textbf {\bibinfo {volume} {407}},\ \bibinfo {pages} {221
  } (\bibinfo {year} {1983}{\natexlab{a}})}\BibitemShut {NoStop}%
\bibitem [{\citenamefont {Hagino}\ and\ \citenamefont
  {Takigawa}(2012)}]{Hag12}%
  \BibitemOpen
  \bibfield  {author} {\bibinfo {author} {\bibfnamefont {K.}~\bibnamefont
  {Hagino}}\ and\ \bibinfo {author} {\bibfnamefont {N.}~\bibnamefont
  {Takigawa}},\ }\Doi {10.1143/PTP.128.1061} {\bibfield  {journal} {\bibinfo
  {journal} {Prog. Theor. Phys.},\ }\textbf {\bibinfo {volume} {128}},\
  \bibinfo {pages} {1061} (\bibinfo {year} {2012})}\BibitemShut {NoStop}%
\bibitem [{\citenamefont {Aguilera}\ \emph {et~al.}(2011)\citenamefont
  {Aguilera} \emph {et~al.}}]{Agu11}%
  \BibitemOpen
  \bibfield  {author} {\bibinfo {author} {\bibfnamefont {E.~F.}\ \bibnamefont
  {Aguilera}} \emph {et~al.},\ }\Doi {10.1103/PhysRevLett.107.092701}
  {\bibfield  {journal} {\bibinfo  {journal} {Phys. Rev. Lett.},\ }\textbf
  {\bibinfo {volume} {107}},\ \bibinfo {pages} {092701} (\bibinfo {year}
  {2011})}\BibitemShut {NoStop}%
\bibitem [{\citenamefont {Aguilera}\ \emph {et~al.}(2009)\citenamefont
  {Aguilera} \emph {et~al.}}]{Agu09}%
  \BibitemOpen
  \bibfield  {author} {\bibinfo {author} {\bibfnamefont {E.~F.}\ \bibnamefont
  {Aguilera}} \emph {et~al.},\ }\Doi {10.1103/PhysRevC.79.021601} {\bibfield
  {journal} {\bibinfo  {journal} {Phys. Rev. C},\ }\textbf {\bibinfo {volume}
  {79}},\ \bibinfo {pages} {021601} (\bibinfo {year} {2009})}\BibitemShut
  {NoStop}%
\bibitem [{\citenamefont {Scuderi}\ \emph {et~al.}(2011)\citenamefont {Scuderi}
  \emph {et~al.}}]{Scu11}%
  \BibitemOpen
  \bibfield  {author} {\bibinfo {author} {\bibfnamefont {V.}~\bibnamefont
  {Scuderi}} \emph {et~al.},\ }\Doi {10.1103/PhysRevC.84.064604} {\bibfield
  {journal} {\bibinfo  {journal} {Phys. Rev. C},\ }\textbf {\bibinfo {volume}
  {84}},\ \bibinfo {pages} {064604} (\bibinfo {year} {2011})}\BibitemShut
  {NoStop}%
\bibitem [{\citenamefont {Hagino}\ \emph {et~al.}(2000)\citenamefont {Hagino},
  \citenamefont {Vitturi}, \citenamefont {Dasso},\ and\ \citenamefont
  {Lenzi}}]{Hag00}%
  \BibitemOpen
  \bibfield  {author} {\bibinfo {author} {\bibfnamefont {K.}~\bibnamefont
  {Hagino}}, \bibinfo {author} {\bibfnamefont {A.}~\bibnamefont {Vitturi}},
  \bibinfo {author} {\bibfnamefont {C.~H.}\ \bibnamefont {Dasso}}, \ and\
  \bibinfo {author} {\bibfnamefont {S.~M.}\ \bibnamefont {Lenzi}},\ }\Doi
  {10.1103/PhysRevC.61.037602} {\bibfield  {journal} {\bibinfo  {journal}
  {Phys. Rev. C},\ }\textbf {\bibinfo {volume} {61}},\ \bibinfo {pages}
  {037602} (\bibinfo {year} {2000})}\BibitemShut {NoStop}%
\bibitem [{\citenamefont {Vinodkumar}\ \emph {et~al.}(2013)\citenamefont
  {Vinodkumar}, \citenamefont {Loveland}, \citenamefont {Yanez}, \citenamefont
  {Leonard}, \citenamefont {Yao}, \citenamefont {Bricault}, \citenamefont
  {Dombsky}, \citenamefont {Kunz}, \citenamefont {Lassen}, \citenamefont
  {Morton}, \citenamefont {Ottewell}, \citenamefont {Preddy},\ and\
  \citenamefont {Trinczek}}]{Vin13}%
  \BibitemOpen
  \bibfield  {author} {\bibinfo {author} {\bibfnamefont {A.~M.}\ \bibnamefont
  {Vinodkumar}}, \bibinfo {author} {\bibfnamefont {W.}~\bibnamefont
  {Loveland}}, \bibinfo {author} {\bibfnamefont {R.}~\bibnamefont {Yanez}},
  \bibinfo {author} {\bibfnamefont {M.}~\bibnamefont {Leonard}}, \bibinfo
  {author} {\bibfnamefont {L.}~\bibnamefont {Yao}}, \bibinfo {author}
  {\bibfnamefont {P.}~\bibnamefont {Bricault}}, \bibinfo {author}
  {\bibfnamefont {M.}~\bibnamefont {Dombsky}}, \bibinfo {author} {\bibfnamefont
  {P.}~\bibnamefont {Kunz}}, \bibinfo {author} {\bibfnamefont {J.}~\bibnamefont
  {Lassen}}, \bibinfo {author} {\bibfnamefont {A.~C.}\ \bibnamefont {Morton}},
  \bibinfo {author} {\bibfnamefont {D.}~\bibnamefont {Ottewell}}, \bibinfo
  {author} {\bibfnamefont {D.}~\bibnamefont {Preddy}}, \ and\ \bibinfo {author}
  {\bibfnamefont {M.}~\bibnamefont {Trinczek}},\ }\Doi
  {10.1103/PhysRevC.87.044603} {\bibfield  {journal} {\bibinfo  {journal}
  {Phys. Rev. C},\ }\textbf {\bibinfo {volume} {87}},\ \bibinfo {pages}
  {044603} (\bibinfo {year} {2013})}\BibitemShut {NoStop}%
\bibitem [{\citenamefont {G\'omez~Camacho}\ \emph {et~al.}(2011)\citenamefont
  {G\'omez~Camacho}, \citenamefont {Aguilera}, \citenamefont {Gomes},\ and\
  \citenamefont {Lubian}}]{Gom11}%
  \BibitemOpen
  \bibfield  {author} {\bibinfo {author} {\bibfnamefont {A.}~\bibnamefont
  {G\'omez~Camacho}}, \bibinfo {author} {\bibfnamefont {E.~F.}\ \bibnamefont
  {Aguilera}}, \bibinfo {author} {\bibfnamefont {P.~R.~S.}\ \bibnamefont
  {Gomes}}, \ and\ \bibinfo {author} {\bibfnamefont {J.}~\bibnamefont
  {Lubian}},\ }\Doi {10.1103/PhysRevC.84.034615} {\bibfield  {journal}
  {\bibinfo  {journal} {Phys. Rev. C},\ }\textbf {\bibinfo {volume} {84}},\
  \bibinfo {pages} {034615} (\bibinfo {year} {2011})}\BibitemShut {NoStop}%
\bibitem [{\citenamefont {Nakatsukasa}\ \emph {et~al.}(2004)\citenamefont
  {Nakatsukasa}, \citenamefont {Yabana}, \citenamefont {Ito}, \citenamefont
  {Kobayashi},\ and\ \citenamefont {Ueda}}]{Nak04}%
  \BibitemOpen
  \bibfield  {author} {\bibinfo {author} {\bibfnamefont {T.}~\bibnamefont
  {Nakatsukasa}}, \bibinfo {author} {\bibfnamefont {K.}~\bibnamefont {Yabana}},
  \bibinfo {author} {\bibfnamefont {M.}~\bibnamefont {Ito}}, \bibinfo {author}
  {\bibfnamefont {M.}~\bibnamefont {Kobayashi}}, \ and\ \bibinfo {author}
  {\bibfnamefont {M.}~\bibnamefont {Ueda}},\ }\Doi {10.1143/PTPS.154.85}
  {\bibfield  {journal} {\bibinfo  {journal} {Prog. Theor. Phys. Suppl.},\
  }\textbf {\bibinfo {volume} {154}},\ \bibinfo {pages} {85} (\bibinfo {year}
  {2004})}\BibitemShut {NoStop}%
\bibitem [{\citenamefont {Broglia}\ and\ \citenamefont {Winther}(1991)}]{BW}%
  \BibitemOpen
  \bibfield  {author} {\bibinfo {author} {\bibfnamefont {R.~A.}\ \bibnamefont
  {Broglia}}\ and\ \bibinfo {author} {\bibfnamefont {A.}~\bibnamefont
  {Winther}},\ }\href@noop {} {\emph {\bibinfo {title} {Heavy Ion
  Reactions}}},\ Frontiers in Physics\ (\bibinfo  {publisher}
  {Addison-Wesley},\ \bibinfo {year} {1991})\BibitemShut {NoStop}%
\bibitem [{\citenamefont {Austern}\ \emph {et~al.}(1987)\citenamefont
  {Austern}, \citenamefont {Iseri}, \citenamefont {Kamimura}, \citenamefont
  {Kawai}, \citenamefont {Rawitscher},\ and\ \citenamefont {Yahiro}}]{Aus87}%
  \BibitemOpen
  \bibfield  {author} {\bibinfo {author} {\bibfnamefont {N.}~\bibnamefont
  {Austern}}, \bibinfo {author} {\bibfnamefont {Y.}~\bibnamefont {Iseri}},
  \bibinfo {author} {\bibfnamefont {M.}~\bibnamefont {Kamimura}}, \bibinfo
  {author} {\bibfnamefont {M.}~\bibnamefont {Kawai}}, \bibinfo {author}
  {\bibfnamefont {G.}~\bibnamefont {Rawitscher}}, \ and\ \bibinfo {author}
  {\bibfnamefont {M.}~\bibnamefont {Yahiro}},\ }\href@noop {} {\bibfield
  {journal} {\bibinfo  {journal} {Phys. Rep.},\ }\textbf {\bibinfo {volume}
  {154}},\ \bibinfo {pages} {125} (\bibinfo {year} {1987})}\BibitemShut
  {NoStop}%
\bibitem [{\citenamefont {Dasso}\ \emph
  {et~al.}(1983){\natexlab{b}}\citenamefont {Dasso}, \citenamefont {Landowne},\
  and\ \citenamefont {Winther}}]{Das83}%
  \BibitemOpen
  \bibfield  {author} {\bibinfo {author} {\bibfnamefont {C.~H.}\ \bibnamefont
  {Dasso}}, \bibinfo {author} {\bibfnamefont {S.}~\bibnamefont {Landowne}}, \
  and\ \bibinfo {author} {\bibfnamefont {A.}~\bibnamefont {Winther}},\ }\Doi
  {10.1016/0375-9474(83)90578-X} {\bibfield  {journal} {\bibinfo  {journal}
  {Nucl. Phys. A},\ }\textbf {\bibinfo {volume} {405}},\ \bibinfo {pages} {381
  } (\bibinfo {year} {1983}{\natexlab{b}})}\BibitemShut {NoStop}%
\bibitem [{\citenamefont {Dasso}(1997)}]{Das97}%
  \BibitemOpen
  \bibfield  {author} {\bibinfo {author} {\bibfnamefont {C.~H.}\ \bibnamefont
  {Dasso}},\ }\Doi {10.1088/0954-3899/23/10/007} {\bibfield  {journal}
  {\bibinfo  {journal} {Journal of Physics G: Nuclear and Particle Physics},\
  }\textbf {\bibinfo {volume} {23}},\ \bibinfo {pages} {1203} (\bibinfo {year}
  {1997})}\BibitemShut {NoStop}%
\end{thebibliography}%

\end{document}